\documentclass[12pt]{article}
\usepackage{amssymb}

\newcommand{\R}{\mathbb{R}} 
\newcommand{\C}{\mathbb{C}} 
\newcommand{\tr}{\mbox{tr} \,}

\newcommand{\vect}[1]{\mbox{\boldmath $ #1 $}}

\def\be{\begin{equation}}
\def\ee{\end{equation}}

\makeatletter
\@addtoreset{equation}{section}
\makeatother

\begin{document}
\baselineskip 6mm
\begin{flushright}
quant-ph/0406038
\\
June 7, 2004
\end{flushright}
\vspace{5mm}
\begin{center}
{\bf \Large
Exact solutions of the isoholonomic problem and 

the optimal control problem in holonomic quantum computation

}
\vspace{10mm}

{Shogo Tanimura}%
\renewcommand{\thefootnote}{*}\footnote{Corresponding author}%
\setcounter{footnote}{0}%
\renewcommand{\thefootnote}{\arabic{footnote}}\footnote{%
E-mail: tanimura@mech.eng.osaka-cu.ac.jp
}, 
{Mikio Nakahara}\footnote{E-mail: nakahara@math.kindai.ac.jp}, 
{Daisuke Hayashi}%
\footnote{E-mail: daisuke{\_}hayashi{\_}0102@nifty.com}
\footnote{%
Present address: Fuji Photo Film Co. Ltd., Asaka Technology
Development Center, Asaka, Saitama 351-8585, Japan}
\vspace{10mm}

$ {}^1 $
{\it
Graduate School of Engineering, Osaka City University,
Osaka 558-8585, Japan
}

$ {}^2 $
{\it
Department of Physics, Kinki University,
Higashi-Osaka 577-8502, Japan
}

$ {}^3 $
{\it
Department of Engineering Physics and Mechanics,
Kyoto University, 
\\
Kyoto 606-8501, Japan
}
\vspace{10mm} \\
Abstract
\vspace{4mm} \\
\begin{minipage}{120mm}
The isoholonomic problem in a homogeneous bundle 
is formulated and solved exactly.
The problem takes a form of
a boundary value problem of a variational equation.
The solution is applied to the optimal control problem
in holonomic quantum computer.
We provide a prescription to construct an optimal controller
for an arbitrary unitary gate and apply it
to a $ k $-dimensional unitary
gate which operates on an $ N $-dimensional Hilbert space with
$ N \geq 2k $.
Our construction is applied to several important unitary gates such as
the Hadamard gate, the CNOT gate,
and the two-qubit discrete Fourier transformation gate.
Controllers for these gates are explicitly constructed.
\end{minipage}
\vspace{10mm} 
\\
\begin{minipage}{140mm}
\end{minipage}
\end{center}

\newpage
\setcounter{footnote}{0}
\renewcommand{\thefootnote}{\alph{footnote}}
\baselineskip 6mm

\section{Introduction}
In this paper 
we solve the isoholonomic problem in a homogeneous bundle
and apply this result to the optimal control problem
in holonomic quantum computation.
In other words, this paper has two purposes;
first, we solve a mathematical problem
which has been unsolved
for more than a decade since it was initially proposed by
Montgomery \cite{mont}.
Second, we provide a scheme to construct explicitly 
an optimal controller for arbitrary unitary gate 
in holonomic quantum computation \cite{first,pachos1}.

The isoholonomic problem is one of generalizations 
of the isoperimetric problem.
The isoperimetric problem, also known as Dido's problem,
is originally proposed in the context of plane geometry;
what is the shape of a domain with the largest area surrounded
by a string of a fixed length? The solution is a circle.
The isoperimetric problem has a long history and
various generalizations thereof have been proposed.

The isoholonomic problem is formulated as follows.
Assume that we have a principal fiber bundle
$ (P, M, \pi, G) $ with a connection.
The base space $ M $ is assumed to be a Riemannian manifold.
The isoholonomic problem asks to
find the shortest possible piecewise smooth loop in $M$ with
a given base point $ x_0 \in M $, that produces 
a given element $ g_0 $ of the structure group $ G $ 
as its associated holonomy.

Holonomic structures naturally appear in a mechanical system
and have been studied from various interests \cite{Gui}-\cite{Nakamura}.
Montgomery faced this problem 
when physical chemists attempted 
to observe the non-Abelian Berry phase (the Wilczek-Zee holonomy)
\cite{Berry}-\cite{Zee}
by nuclear magnetic resonance (NMR) experiment.
Montgomery \cite{mont} presented various formulations of the problem,
clarified their relations, and gave partial answers.
However, even in such an idealized case like a homogeneous bundle,
it was difficult to obtain a complete solution to the problem,
which remained as an open problem to date.

A decade later after Montgomery's work,
the notion of holonomic quantum computation was proposed by 
Zanardi, Rasetti and Pachos \cite{first, pachos1}, in which
the Wilczek-Zee holonomy is utilized to implement unitary gates necessary
to execute a quantum algorithm.
Since then, a large number of researchers \cite{unanyan}-\cite{pachos3}
have been interested in finding control parameters
that implement a desired gate. Optimization of the control
has been an active area of research in view of the decoherence issue.
The problem to find the optimal control is nothing but a typical
isoholonomic problem
and its solution for an arbitrary gate must be urgently provided.

Let us briefly review the idea of quantum computation.
Quantum computation, roughly speaking, consists of three ingredients:
(1) an $ n $-qubit register to store information,
(2) a unitary matrix $ U \in U(2^n) $ which implements a quantum algorithm, and
(3) measurements to extract information from the register.
In an ordinary implementation of a quantum algorithm, 
we take a system whose Hamiltonian $ H(\lambda) $
depends on external control parameters 
$ \lambda = ( \lambda^1, \dots, \lambda^m ) $.
We then properly arrange the parameter sequence $ \lambda(t) $ 
as a function of time $ t $
so that the desired unitary matrix $ U $ is
generated as a time-evolution operator
\begin{equation}
	U = {\mathcal{T}} 
	\exp 
	\left[ 
		- \frac{i}{\hbar} \int_0^T H(\lambda(t)) dt
	\right],
\end{equation}
where $\mathcal{T}$ stands for the time-ordered product.

Holonomic quantum computing \cite{first}, in contrast, makes use of 
the holonomy associated with a loop $ \lambda (t) $ 
in the parameter space.
It has been demonstrated \cite{ours1}
that an arbitrary unitary matrix can be
implemented as a holonomy by choosing an appropriate loop in the
parameter space.
In fact, there are infinitely many loops that produce a given
unitary matrix. 
Here we consider the isoholonomic problem, 
namely, to find the shortest
possible loop in the parameter space that yields the given holonomy. 
This problem has been already analyzed previously in Ref.\cite{ours2},
where various penalty functions useful for numerical search for the optimal 
loop have been employed. Our strategy
here is purely geometrical in nature and no intense numerical computations
are required. In the previous work \cite{Letter}
we found exact optimal loops to produce several unitary gates.
In the present paper we extend the method of optimal loop construction
to implement arbitrary gates.

This paper is organized as follows.
In Section 2 we briefly review the Wilczek-Zee holonomy
to make this paper self-contained and to establish notation
conventions.
In Section 3 we introduce the geometrical setting for the problem
and use it to formulate the isoholonomic problem 
in a variational form.
We derive the associated Euler-Lagrange equation and solve it explicitly.
The solution thus obtained (\ref{complete solution}),
which we call the horizontal extremal curve,
is one of the main results in the first half of the paper.
The remaining problem is to adjust the solution
to satisfy the boundary conditions, namely
the closed loop condition (\ref{loop condition})
and the holonomy condition (\ref{holonomy condition}).
This problem is solved in Section 4 explicitly and we
obtain a set of equations
(\ref{det0})-(\ref{det3}),
which we call the constructing equations of the controller.
These are the main results of this paper
and are machinery to construct a controller
for an arbitrary unitary gate.
In Section 5
this machinery is applied to several well-known important unitary gates 
to demonstrate its power.
Section 6 is devoted to summary and discussions.

\section{Wilczek-Zee holonomy as a unitary gate}

\subsection{Wilczek-Zee holonomy}

Here we briefly review the Wilczek-Zee (WZ) holonomy \cite{wz} 
associated with an adiabatic change of the control parameters along a loop 
in the control manifold.
We consider a quantum system that has a finite number $ N $ of states.
Let $ \{ H(\lambda) \} $ be a family of Hamiltonians 
parametrized smoothly 
by $ \lambda = ( \lambda^1, \dots, \lambda^m ) $ $ \in M $, 
where the set of control parameters $ M $ is called a control manifold.
Eignevalues and eigenstates of $ H(\lambda) $ are labeled as
\begin{equation}
	H(\lambda) |l, \alpha; \lambda \rangle 
	= \varepsilon_l (\lambda) |l, \alpha; \lambda \rangle
	\qquad 
	( l=1,\ldots,L; \, \alpha = 1, \ldots, k_l ),
\end{equation}
where the $ l $-th eigenvalue
$ \varepsilon_l(\lambda)$ is $ k_l $-fold degenerate.
Assume that no level crossings take place,
namely,
$ \varepsilon_l(\lambda) \ne \varepsilon_{l'} (\lambda) $
for arbitrary $ \lambda $ 
if $ l \ne l' $.
Then it follows that $\sum_{l=1}^L k_l = N $. 
The eigenvectors satisfy the orthonormal condition,
$ \langle l, \alpha ; \lambda | l', \beta ; \lambda \rangle 
= \delta_{ll'}\delta_{\alpha \beta} $.
It is important to note that there is $U( k_l)$ {\it gauge} freedom
in the choice of 
$ \{ |l, \alpha; \lambda \rangle \, | \, \alpha = 1, \dots, k_l \} $ 
at each $\lambda$ and $ l $.
Namely, we may redefine the eigenvectors by any unitary 
matrix $ h \in U (k_l) $ as
\begin{equation}
	|l, \alpha; \lambda \rangle 
	\mapsto
	\sum_{\beta=1}^{k_l} 
	|l, \beta; \lambda \rangle h_{\beta \alpha} (\lambda)
	\label{freedom}
\end{equation}
without violating the orthonormal condition.

We adiabatically change the parameters $ \lambda(t) $ 
as a function of time $ t $
along a closed loop in the control manifold
so that $ \lambda(T) = \lambda(0) $.
It is assumed that the {\it adiabaticity} is satified, namely,
\begin{equation}
	\{ \varepsilon_l (\lambda(t)) - \varepsilon_{l'} (\lambda(t)) \} T
	\gg
	2 \pi \hbar
\end{equation}
is satisfied for $ l \ne l' $ during $ 0 \le t \le T $.
In other words, we change the parameters so slowly 
that no resonant transitions take place
between different energy levels \cite{Messiah}.

We will concentrate exclusively on the ground state of the system
and drop the index $ l \; (=1) $ in the following. 
Accordingly, the basis vectors that span the ground state eigenspace
are written as 
$ | \alpha; \lambda \rangle $, $ ( \alpha = 1, \ldots, k ) $
and arranged in an $ N \times k $ matrix form as
\begin{equation}
	V(\lambda) =
	\Big(
		| 1; \lambda \rangle, 
		| 2; \lambda \rangle, 
		\dots,
		| k; \lambda \rangle 
	\Big),
	\label{frame}
\end{equation}
which is called an orthonormal $ k $-frame at $ \lambda \in M $.
%
The system evolves, with a given $\lambda (t)$, according to the Schr\"odinger equation
\begin{equation}
	i \hbar \frac{d}{dt} |\psi_{\alpha}(t) \rangle
	= H(\lambda(t)) |\psi_{\alpha}(t) \rangle.
	\label{eq:1}
\end{equation}
Suppose the initial condition is $\lambda (0)=\lambda_0$ and
$ |\psi_{\alpha}(0) \rangle = |\alpha; \lambda_0 \rangle $.
The adiabatic theorem \cite{Messiah} tells us that
the state $ |\psi_{\alpha}(t) \rangle $ remains in the ground state 
eigenspace during the time-evolution.
Therefore $ |\psi_{\alpha}(t) \rangle $ is expanded as
\begin{equation}
	|\psi_{\alpha}(t) \rangle 
	= \sum_{\beta = 1}^k |\beta; \lambda(t) \rangle
	\, c_{\beta \alpha}(t).
	\label{eq:sol}
\end{equation}
By substituting (\ref{eq:sol}) into (\ref{eq:1}), we find
\begin{equation}
	\frac{d}{dt} c_{\beta \alpha} (t)
	= - \frac{i}{\hbar} \varepsilon(\gamma(t)) c_{\beta \alpha} (t)
	- \sum_{\gamma=1}^k
	\left\langle 
		\beta ; \lambda(t) 
		\left| \frac{d}{dt} \right| \gamma; \lambda(t) 
	\right\rangle c_{\gamma \alpha}(t),
\end{equation}
whose formal solution is 
\begin{equation}
	c_{\beta \alpha} (t) = 
	\exp \! \left(
		- \frac{i}{\hbar} \int_0^t \varepsilon(s) ds
	\right)
	{\mathcal{T}} 
	\exp \! \left(-\int_0^t {\cal A}(s) ds \right)_{\! \beta \alpha}
	\label{c}
\end{equation}
with the matrix-valued function
\begin{equation}
	{\cal A}_{\beta \alpha}(t) 
	=
	\left\langle \beta ; \lambda(t) 
		\left| \frac{d}{dt} \right|
	\alpha; \lambda(t) \right\rangle
	=
	\sum_{\mu=1}^k
	\left\langle \beta ; \lambda
		\left| \frac{\partial}{\partial \lambda^\mu } \right|
	\alpha; \lambda \right\rangle
	\frac{d \lambda^\mu}{dt}.
\end{equation}
It is easily verified that
$ {\cal A}_{\beta \alpha}^* = - {\cal A}_{\alpha \beta} $
since $ \{|\alpha; \lambda(t) \rangle\} $ is orthonormal.
We introduce a $ \mathfrak{u}(k) $-valued one-form\footnote{
We denote the Lie algebra of the Lie group $ U(k) $ 
by $ \mathfrak{u}(k) $,
which is the set of $ k $-dimensional skew-Hermite matrices.
}
\begin{equation}
	{\cal A}_{\beta \alpha} (\lambda) = 
	\sum_{\mu=1}^k
	\left\langle \beta ; \lambda
		\left| \frac{\partial}{\partial \lambda^{\mu}} \right|
	\alpha; \lambda \right\rangle d\lambda^{\mu},
	\label{WZ conn}
\end{equation}
which is called the Wilczek-Zee (WZ) connection.
Then the unitary matrix appearing in (\ref{c}) is rewritten as
\begin{equation}
	{\mit\Gamma}(t) = 
	{\cal{P}} \exp
	\left(
		-\int_{\lambda(0)}^{\lambda(t)} {\mathcal{A}}
	\right),
\end{equation}
where $\mathcal{P}$ stands for the path-ordered product.
As noted in (\ref{freedom}) the frame (\ref{frame}) can be redefined by
a family of unitary matrices $ h(\lambda) \in U(k) $.
The WZ connection transforms under the change of frame as
\begin{equation}
	{\cal A} \mapsto 
	{\cal A}' = h^{\dagger} {\mathcal{A}} h + h^{\dagger} d h.
\end{equation}
This is nothing but the gauge transformation rule of a non-Abelian
gauge potential \cite{gtp}. 

We assumed that the control parameter $ \lambda(t) $ comes back to
the initial point $ \lambda(T) = \lambda(0) = \lambda_0 $.
However, the state $ | \psi_{\alpha}(T) \rangle $ fails to assume 
the initial state and is subject to a unitary rotation as
\begin{equation}
	| \psi_{\alpha}(T) \rangle 
	= 
	\exp \! \left(
		- \frac{i}{\hbar} \int_0^T \varepsilon(s) ds
	\right)
	\sum_{\beta = 1}^k |\psi_{\beta}(0) \rangle
	{\mit\Gamma}_{\beta \alpha} (T).
	\label{final state}
\end{equation}
The unitary matrix
\begin{equation}
	{\mit\Gamma} [{\lambda}] := {\mit\Gamma}(T) 
	= {\mathcal{P}} \exp
	\left( -\oint_{\lambda} {\mathcal{A}} \right) \in U(k)
	\label{hol}
\end{equation}
is called the holonomy matrix associated with the loop $ \lambda(t) $.
It is important to realize that $ {\mit\Gamma}[\lambda] $ is independent of the
parametrization of the loop $ \lambda(t) $, 
namely, it is independent of how fast the loop 
$\lambda$ is traversed, so long as the
adiabaticity is observed, and that it depends only
on the geometrical image of $ \lambda $ in $ M $.

\subsection{Quantum computation with holonomy}

In quantum computation one implements a quantum algorithm by a product of
various unitary gates.
It is a natural idea to use the WZ holonomy to produce unitary gates
necessary for quantum computation. 
Zanardi and Rasetti \cite{first} were the first who 
proposed this holonomic quantum computation (HQC).
To implement an $ n $-qubit resistor we take
a quantum system whose ground state is $ k $-fold degenerate where
$ k = 2^n $. We call the $ N $-dimensional Hilbert space a working space
and call the $ k $-dimensional subspace a qubit space.
Then by changing the control parameter adiabatically
we will obtain any unitary gate 
as a resultant holonomy (\ref{hol}).
Of course we need to design an appropriate control loop $ \lambda $
to implement a particular unitary gate.
It is easy, in principle, to compute the holonomy for a given loop. 
In contrast, to find a loop $\lambda$
which produces a specified unitary matrix $ {\mit\Gamma} $ as its holonomy
is far from trivial. 
Moreover, to build a working quantum computer
it is strongly desired to reduce the time required to manipulate the computer
since a sequence of operations should be carried out 
before decoherence extinguishes quantum information from the system.
At the same time, the control parameter must be changed as slowly as possible
to keep adiabaticity intact. Therefore our task
is to find a control loop as short as possible to fulfill these seemingly
opposed conditions. This is a typical example of 
the so-called isoholonomic problem,
which is first formulated by Montgomery \cite{mont}.
In the next section we introduce a geometric setting 
in a form suitable for our expositions.

\section{Formulation of the problem and its solution}

\subsection{Geometrical setting}

The WZ connection is identified with 
the canonical connection \cite{KN} 
of the homogenous bundle,
as pointed out by Fujii \cite{fujii1}.
While precise definitions of these terms can be found in books \cite{gtp, KN}, 
we outline the geometrical setting of the
problem here to make this paper self-contained.

Suppose that the system 
has a family of Hamiltonians acting on the Hilbert space 
$ {\mathbb{C}}^N $
and that the ground state of each Hamiltonian is
$ k $-fold degenerate $ ( k < N ) $.
The most natural mathematical setting to describe this system
is the principal bundle
$ (S_{N, k}(\mathbb{C}), $ $G_{N, k}(\mathbb{C}), \pi, U(k)) $,
which consists of
the Stiefel manifold $ S_{N, k}(\mathbb{C}) $, 
the Grassmann manifolds $ G_{N, k}(\mathbb{C})$, 
the projection map $ \pi : S_{N, k}(\mathbb{C}) \to G_{N, k}(\mathbb{C})$,
and the unitary group $ U(k) $
as explained below.

The Stiefel manifold is the set of orthonormal $k$-frames in ${\mathbb{C}}^N$,
\begin{equation}
	S_{N, k} ( {\mathbb{C}} ) =
	\{ V \in M(N, k; \mathbb{C}) \, | \, V^{\dagger} V = I_k\},
\end{equation}
where $ M(N, k; \mathbb{C}) $ is the set of $N \times k$ complex matrices
and $I_k$ is the $ k $-dimensional unit matrix. 
The unitary group $ U(k) $ acts on $ S_{N, k} ({\mathbb{C}}) $ from the right
\begin{equation}
	S_{N, k}({\mathbb{C}}) \times U(k)
	\to
	S_{N, k}({\mathbb{C}}),
	\qquad
	(V, h) \mapsto Vh
\end{equation}
by means of matrix product.
It should be noted that this action is free.
In other words, $ h = I_k $
if there exists a point $V \in S_{N, k}({\mathbb{C}})$ such that
$ Vh = V $.

The Grassmann manifold is defined as the set
of $ k $-dimensional hyperplanes in ${\mathbb{C}}^N$,
\begin{eqnarray}
	G_{N, k}(\mathbb{C})
	= \{ P \in  M(N,N;\mathbb{C}) \, | \, 
	P^2=P, \, P^{\dagger} = P, \, \mathrm{tr} P=k \},
\end{eqnarray}
where $ P $ is a projection operator to a hyperplane in $\mathbb{C}^N$
and the condition $\mathrm{tr}P=k$ guarantees that the hyperplane is
indeed $k$-dimensional.

The projection map
$ \pi : S_{N, k}({\mathbb{C}})$ $ \to G_{N, k}({\mathbb{C}})$
is defined as
\begin{equation}
	\pi: V \mapsto P := VV^{\dagger}.
\end{equation}
It is easily proved that the map $ \pi $ is surjective.
Namely, for any $ P \in G_{N, k}({\mathbb{C}}) $,
there is $ V \in S_{N, k}({\mathbb{C}}) $ such that $ \pi(V) = P $.
The right action of $ h \in U(k) $ sends 
a point $ V \in S_{N, k}({\mathbb{C}})$ to 
a point $ Vh $ on the same fiber since
\begin{equation}
	\pi(Vh) 
	= (Vh)(Vh)^{\dagger} 
	= V hh^{\dagger} V^{\dagger} 
	= VV^{\dagger} 
	= \pi(V).
\end{equation}
Thus the Stiefel manifold $ S_{N, k}({\mathbb{C}})$ becomes 
a principal bundle over $ G_{N, k}({\mathbb{C}}) $
with the structure group $ U(k) $.

Moreover, the group $U(N)$ acts on both $S_{N, k}({\mathbb{C}})$ and
$G_{N, k}({\mathbb{C}})$ as
\begin{eqnarray}
&&	U(N) \times S_{N, k}({\mathbb{C}}) \to S_{N, k}({\mathbb{C}}),
	\qquad
	(g, V) \mapsto gV,
	\label{left action}
\\
&&	U(N) \times G_{N, k}({\mathbb{C}}) \to G_{N, k}({\mathbb{C}}),
	\qquad
	(g, P) \mapsto g P g^{\dagger}
\end{eqnarray}
by matrix product. It is easily verified that
$ \pi(gV) = g \pi(V) g^{\dagger} $.
This action is transitive, namely, 
there is $ g \in U(N) $ for any $ V, V' \in S_{N,k} ({\mathbb{C}}) $ 
such that $ V' = g V $.
There is also $ g \in U(N) $ for any $ P, P' \in G_{N,k} ({\mathbb{C}}) $ 
such that $ P' = g P g^\dagger $.
The stabilizer group of each point in $ S_{N,k} ({\mathbb{C}}) $ is
isomorphic to $ U(N-k) $ while
that of each point in $ G_{N,k} ({\mathbb{C}}) $ is
isomorphic to $ U(k) \times U(N-k) $.
Thus, they are homogeneous spaces and the fiber bundle
\begin{equation}
	\pi : 
	S_{N, k}({\mathbb{C}}) \cong U(N)/U(N-k)
	\to 
	G_{N, k}({\mathbb{C}}) \cong U(N)/(U(k) \times U(N-k))
\end{equation}
is call a homogeneous bundle.

The canonical connection form on $ S_{N, k}({\mathbb{C}}) $ 
is defined as a $ \mathfrak{u}(k)$-valued one-form 
\begin{equation}
	A = V^{\dagger} d V,
\end{equation}
which is a generalization of the WZ connection (\ref{WZ conn}).
This is characterized as the unique connection 
that is invariant under the action (\ref{left action}).
The associated curvature two-form is then defined as
\begin{equation}
	F
=	d A + A \wedge A 
=	dV^{\dagger} \wedge dV + V^{\dagger} dV \wedge V^{\dagger} d V
=	dV^{\dagger} \wedge (I_N - VV^{\dagger})dV.
\end{equation}

These manifolds are equipped with Riemannian metrics. 
We define a metric
\begin{equation}
	\| dV \|^2 = {\mathrm{tr}} \, ( dV^{\dagger} dV )
\end{equation}
for the Stiefel manifold and
\begin{equation}
	\| dP \|^2 = {\mathrm{tr}} \, (dP dP)
\end{equation}
for the Grassmann manifold.

\subsection{The isoholonomic problem}

Here we reformulate the WZ holonomy 
in terms of the geometric terminology introduced above.
The state vector $ \psi(t) \in \C^N $ evolves 
according to the Schr{\"o}dinger equation
\begin{equation}
	i \hbar \frac{d}{dt} \psi(t) = H(t) \psi(t).
	\label{Schrodinger}
\end{equation}
The Hamiltonian admits a spectral decomposition
\begin{equation}
	H(t) = \sum_{l=1}^L \varepsilon_l (t) P_l (t)
	\label{decomposition of H}
\end{equation}
with projection operators $ P_l (t) $.
Therefore, the set of energy eigenvalues 
$ ( \varepsilon_1, \dots, \varepsilon_L ) $
and orthogonal projectors
$ ( P_1, \dots, P_L ) $
constitutes a complete set of control parameters of the system.
Now we concentrate on the eigenspace associated with the lowest energy,
which is assumed to be identically zero, $ \varepsilon_1 \equiv 0 $.
We write $ P_1(t) $ as $ P(t) $ for simplicity.
Suppose that the degree of degeneracy 
$ k = \tr P(t) $ is constant.
For each $ t $, there exists $ V(t) \in S_{N,k} (\C) $ such that
$ P(t) = V(t) V^\dagger (t) $.
By adiabatic approximation we mean substitution of $ \psi(t) \in \C^N $
by a reduced state vector $ \phi(t) \in \C^k $ as
\begin{equation}
	\psi(t) = V(t) \phi(t).
	\label{adiabatic}
\end{equation}
Since $ H(t) \psi(t) = \varepsilon_1 \psi(t) = 0 $,
the Schr{\"o}dinger equation (\ref{Schrodinger}) becomes
\begin{equation}
	\frac{d \phi}{dt} 
	+ V^\dagger \frac{dV}{dt} \phi(t) = 0
	\label{Schrodinger V phi}
\end{equation}
and its formal solution is written as
\begin{equation}
	\phi(t) =
	{\cal P}
	\exp \!
	\left(
		- \int V^\dagger dV
	\right)
	\phi(0).
	\label{formal solution phi}
\end{equation}
Therefore $ \psi(t) $ is written as
\begin{equation}
	\psi(t) =
	V(t)
	{\cal P}
	\exp \!
	\left(
		- \int V^\dagger dV
	\right)
	V^\dagger (0)
	\psi(0).
	\label{formal solution psi}
\end{equation}
In particular, 
when the control parameter comes back to the initial point as $ P(T) = P(0) $,
the holonomy $ {\mit\Gamma} \in U(k) $ is defined via\footnote{%
The definition of the holonomy presented here is slightly different
from the one given in the previous Letter \cite{Letter}.
To make a correct sense as a unitary gate the holonomy is to be defined 
in the present form.
}
\begin{equation}
	\psi(T) = V(0) {\mit\Gamma} \, \phi(0)
	\label{def Gamma}
\end{equation}
and it is given explicitly as
\begin{equation}
	{\mit\Gamma} = V(0)^\dagger \, V(T) \,
	{\cal P}
	\exp \!
	\left(
		- \int V^\dagger dV
	\right).
	\label{Gamma}
\end{equation}
If the condition
\begin{equation}
	V^\dagger \frac{dV}{dt} = 0
	\label{horizontal}
\end{equation}
is satisfied, the curve $ V(t) $ in $ S_{N,k} (\C) $
is called a horizontal lift of
the curve $ P(t) = \pi( V(t) ) $ in $ G_{N,k} (\C) $.
Then the holonomy (\ref{Gamma}) is reduced to
\begin{equation}
	{\mit\Gamma} 
	= V^\dagger (0) V(T) \in U(k).
	\label{hol time evol}
\end{equation}

Now we are ready to state the isoholonomic problem
in the present context;
given a specified unitary gate $ U_{\rm gate} \in U(k) $
and a fixed point $ P_0 \in G_{N,k} (\C) $,
find the shortest loop $ P(t) $ in $ G_{N,k} (\C) $ 
with the base points $ P(0) = P(T) = P_0 $
whose horizontal lift $ V(t) $ in $ S_{N,k} (\C) $ produces a holonomy
$ {\mit\Gamma} $ that coincides with $ U_{\rm gate} $.
This problem was first motivated from experimental study of geometric phase and
investigated in detail from a mathematician's viewpoint 
by Montgomery \cite{mont}.

We now formulate the isoholonomic problem as a variational problem.
The length of the horizontal curve $ V(t) $ is evaluated by the functional
\begin{equation}
	S[V, {\mit\Omega}] = \int_0^T
	\left\{
		{\rm tr} \!
		\left( \frac{dV^\dagger}{dt} \frac{dV}{dt} \right)
		-
		{\rm tr} \! 
		\left( {\mit\Omega} \, V^\dagger \frac{dV}{dt} \right)
	\right\} dt,
	\label{S}
\end{equation}
where $ {\mit\Omega} (t) \in \mathfrak{u} (k) $ is a Lagrange multiplier
to impose the horizontal condition (\ref{horizontal}) on the curve $ V(t) $.
Note that the value of the functional $ S $ is equal to
the length of the projected curve $ P(t)= \pi(V(t)) $,
\begin{equation}
	S = \int_0^T
		\frac{1}{2}
		{\rm tr} \!
		\left( \frac{dP}{dt} \frac{dP}{dt} \right)
	dt.
	\label{length of P}
\end{equation}
Thus the problem is formulated as follows;
find a curve $ V(t) $ 
that attains an extremal value of the functional (\ref{S})
and satisfies the boundary condition (\ref{hol time evol}).

\subsection{The solution: horizontal extremal curve}

Our task is to find a solution 
of the variational problem of the functional (\ref{S}).
Now we derive the associated Euler-Lagrange equation
and solve it explicitly.
A variation of the curve $ V(t) $ is defined by
an arbitrary smooth function $ \eta(t) \in \mathfrak{u}(N) $ 
such that $ \eta(0) = \eta(T) = 0 $
and an infinitesimal parameter $ \epsilon \in \R $ as
\begin{equation}
	V_\epsilon (t) = 
	( 1 + \epsilon \eta(t) ) V(t).
\end{equation}
By substituting $ V_\epsilon (t) $ into (\ref{S}) and
differentiating with respect to $ \epsilon $, 
the extremal condition yields
\begin{eqnarray}
	0 = 
	\left. \frac{d S}{d \epsilon} \right|_{\epsilon = 0}
&=&	
	\int_0^T
	\tr \Big\{
		\dot{\eta} 
		( V \dot{V}^\dagger 
		- \dot{V} V^\dagger 
		- V {\mit\Omega} V^\dagger )
	\Big\} \, dt
	\nonumber \\
&=&	\bigg[
	\tr \big\{
		\eta
		( V \dot{V}^\dagger 
		- \dot{V} V^\dagger 
		- V {\mit\Omega} V^\dagger )
	\big\} 
	\bigg]_{t=0}^{t=T}
	\nonumber \\
&&	-
	\int_0^1
	\tr \Big\{
		\eta \frac{d}{dt}
		( V \dot{V}^\dagger 
		- \dot{V} V^\dagger 
		- V {\mit\Omega} V^\dagger )
	\Big\} \, dt.
\end{eqnarray}
Thus we obtain the Euler-Lagrange equation
\begin{equation}
		\frac{d}{dt}
		( \dot{V} V^\dagger - V \dot{V}^\dagger 
		+ V {\mit\Omega} V^\dagger )
		= 0.
		\label{extremal}
\end{equation}
We reproduce the horizontal equation
$ V^\dagger  \dot{V} = 0 $ from the extremal condition with respect to
$ {\mit\Omega}(t) $. 
Finally, the isoholonomic problem is reduced to the set of equations
(\ref{horizontal}) and (\ref{extremal}),
which we call a horizontal extremal equation.
It may be regarded 
as a homogeneous-space version of the Wong equation \cite{Wong}.

Next, we solve the equations (\ref{horizontal}) and (\ref{extremal}).
The equation (\ref{extremal}) is integrated to yield
\begin{equation}
		\dot{V} V^\dagger - V \dot{V}^\dagger
		+ V {\mit\Omega} V^\dagger 
		= \mbox{const}
		= X \in \mathfrak{u}(N).
		\label{extremal integrated}
\end{equation}
Conjugation of the horizontal condition (\ref{horizontal}) yields
$ \dot{V}^\dagger V = 0 $.
Then, by multiplying $ V $ on (\ref{extremal integrated}) from the right
we obtain
\begin{equation}
	\dot{V} + V {\mit\Omega} = X V.
	\label{extremal integrated 2}
\end{equation}
By multiplying $ V^\dagger $ on (\ref{extremal integrated 2}) from the left
we obtain
\begin{equation}
	{\mit\Omega} 
	= V^\dagger X V.
	\label{extremal integrated 3}
\end{equation}
The equation (\ref{extremal integrated 2}) implies
$ \dot{V} = X V - V {\mit\Omega} $, 
and hence the time derivative of $ {\mit\Omega}(t) $ becomes
\begin{equation}
	\dot{{\mit\Omega}}
	= V^\dagger X \dot{V} + \dot{V}^\dagger X V
	= V^\dagger X ( X V - V {\mit\Omega} )
	+ ( - V^\dagger X + {\mit\Omega} V^\dagger ) X V
	= [ {\mit\Omega}, {\mit\Omega} ] = 0.
	\label{extremal integrated 4}
\end{equation}
Therefore, $ {\mit\Omega}(t) $ is actually a constant.
Thus the solution of 
(\ref{extremal integrated 2}) and (\ref{extremal integrated 3}) is
\begin{equation}
	V(t) = e^{t X} \, V_0 \, e^{-t {\mit\Omega}},
	\qquad
	{\mit\Omega} = V_0^\dagger X V_0.
	\label{extremal solution}
\end{equation}
We call this solution the horizontal extremal curve.
Then (\ref{extremal integrated}) becomes
$$
	( XV - V {\mit\Omega} ) V^\dagger
	- V ( - V^\dagger X + {\mit\Omega} V^\dagger )
	+ V {\mit\Omega} V^\dagger 
	= X,
$$
which is arranged as
\begin{equation}
	X -
	( V V^\dagger X 
	+ X V V^\dagger
	- V V^\dagger X V V^\dagger )
	= 0,
	\label{constraint}
\end{equation}
where we used (\ref{extremal integrated 3}). We may take, without loss of
generality,
\begin{equation}
	V_0 = 
	\left(
		\begin{array}{c}
		I_k \\ 0
		\end{array}
	\right)
	\in S_{N,k} (\C)
	\label{V_0}
\end{equation}
as the initial point.
We can parametrize $ X \in \mathfrak{u}(N) $, 
which satisfies (\ref{extremal integrated 3}), as
\begin{equation}
	X = 
	\left(
		\begin{array}{cc}
		{\mit\Omega}   & W \\
		-W^\dagger & Z 
		\end{array}
	\right)
	\label{X}
\end{equation}
with $ W \in M(k,N-k;\C) $ and $ Z \in \mathfrak{u} (N-k) $.
Then the constraint equation 
(\ref{constraint}) forces us to choose 
\begin{equation}
	Z = 0.
	\label{Z}
\end{equation}

Finally, we obtained a complete set of solution
(\ref{extremal solution})
of the horizontal extremal equation (\ref{horizontal}) and (\ref{extremal}).
When we take the initial point $ V_0 $ as (\ref{V_0}),
the solutions are parametrized by constant matrices
$ {\mit\Omega} \in \mathfrak{u}(k) $ and $ W \in M(k,N-k;\C) $.
For definiteness we write down the complete solution
\begin{equation}
	V(t) = e^{t X} \, V_0 \, e^{-t {\mit\Omega}},
	\qquad
	X = 
	\left(
		\begin{array}{cc}
		{\mit\Omega}   & W \\
		-W^\dagger & 0
		\end{array}
	\right).
	\label{complete solution}
\end{equation}
This is one of our main results.
We call the matrix $ X $ 
a controller.
At this time the holonomy (\ref{hol time evol}) is expressed as
\begin{equation}
	{\mit\Gamma}
	= V^\dagger (0) V(T) 
	= V^\dagger_0 \, e^{T X} \, V_0 \, e^{-T {\mit\Omega}}
	\in U(k).
	\label{holonomy}
\end{equation}
These results (\ref{complete solution}) and (\ref{holonomy})
have been also given in Montgomery's paper%
\footnote{
In his paper \cite{mont}
Montgomery cited B{\"a}r's theorem to complete the proof. 
However, B{\"a}r's paper being a diploma thesis,
it is not widely available. 
Therefore we took a more direct approach to justify them.
}.
In the present paper we took a different approach from his.
Here we wrote down the Euler-Lagrange equation and solved it directly.

We evaluate the length of the extremal curve for later convenience
by substituting (\ref{complete solution}) into (\ref{length of P}) as
\begin{equation}
	S = \int_0^T
		\frac{1}{2}
		{\rm tr} \!
		\left( \frac{dP}{dt} \frac{dP}{dt} \right)
	dt
	= \tr ( W^\dagger W ) \, T.
	\label{length}
\end{equation}

\section{Solution to the inverse problem}

Once the solution (\ref{complete solution})
of the horizontal extremal equation is obtained,
the remaining problem is to find the matrices
$ {\mit\Omega} $ and $ W $
that satisfy the closed loop condition
\begin{equation}
	V(T) V^\dagger (T) 
	= e^{TX} V_0 V_0^\dagger e^{-TX} 
	= V_0 V_0^\dagger
	\label{loop condition}
\end{equation}
and the holonomy condition
\begin{equation}
	V^\dagger_0 \, V(T)
	= V^\dagger_0 \, e^{T X} \, V_0 \, e^{-T {\mit\Omega}} 
	= U_{\rm gate}
	\label{holonomy condition}
\end{equation}
for a specific unitary gate $ U_{\rm gate} \in U(k) $.
Montgomery \cite{mont} presented this inverse problem as an open problem.
In this section we give a scheme to construct systematically
a series of solutions to this problem
and in the next section we will apply it to
implement various important unitary gates.

\subsection{Equivalence class}

There is a class of equivalent solutions with a given initial condition 
$ V_0 $ and a given final condition $V(T)=V_0 U_{\rm gate} $.
Here we clarify the equivalence relation among solutions $ \{ V(t) \} $ 
that have the form (\ref{complete solution})
and satisfy (\ref{loop condition}) and (\ref{holonomy condition}).

We say that two solutions $ V(t) $ and $ V'(t) $ are equivalent if
there are elements $ g \in U(N) $ and $ h \in U(k) $ such that
$V(t)$ and
\begin{equation}
 V'(t)=	g V(t) h^\dagger.
	\label{equivariance}
\end{equation}
satisfy the same boundary conditions
\begin{equation}
	g V_0 h^\dagger = V_0
	\label{eq:ini}
\end{equation}
and
\begin{equation}
	h U_{\rm gate} h^\dagger = U_{\rm gate}.
	\label{gate condition}
\end{equation}
For the initial point (\ref{V_0}), the condition (\ref{eq:ini})
states that $ g \in U(N) $ must have a block-diagonal form
\begin{equation}
	g = 
	\left(
		\begin{array}{cc}
		h_1 & 0 \\
		0   & h_2
		\end{array}
	\right),
	\qquad
	h = h_1 \in U(k), \quad h_2 \in U(N-k).
\end{equation}
The controller $ X' $ of $V'(t)$ are then
found from
\begin{equation}
V'(t)=	g V(t) h^\dagger 
	= g e^{tX} g^\dagger g V_0 h^\dagger h e^{-t {\mit\Omega}} h^\dagger
	= e^{t g X g^\dagger} g V_0 h^\dagger e^{-t h {\mit\Omega} h^\dagger}
	= e^{t g X g^\dagger} V_0 \, e^{-t h {\mit\Omega} h^\dagger}.
\end{equation}
In summary, two controllers $ X $ and $ X' $ are equivalent if and only if
there are unitary matrices
$ h_1 \in U(k) $ and $ h_2 \in U(N-k) $ such that
\begin{equation}
	X = 
	\left(
		\begin{array}{cc}
		{\mit\Omega}   & W \\
		-W^\dagger & 0
		\end{array}
	\right),
	\quad
	X' = 
	\left(
		\begin{array}{cc}
		h_1 {\mit\Omega} h_1^\dagger & h_1 W h_2^\dagger \\
		- h_2 W^\dagger h_1^\dagger & 0
		\end{array}
	\right),
	\quad
	h_1 U_{\rm gate} h_1^\dagger = U_{\rm gate}.
	\label{equivalent X}
\end{equation}

\subsection{$U(1)$ holonomy}
Here we calculate the holonomy for the case $ N=2 $ and $ k=1 $.
In this case the homogeneous bundle 
$ \pi : S_{2,1} (\C) \to G_{2,1} (\C) $
is the Hopf bundle
$ \pi : S^3 \to S^2 $ with the structure group $ U(1) $
and the WZ holonomy reduces to the Berry phase.
In the subsequent subsection we will generalize this result to
a non-Abelian holonomy.
We normalize the cycle time as $ T=1 $ in the following.
Using real numbers $ w_1, w_2, w_3 \in \R $
we parametrize the controller as
\begin{equation}
	X 
=
	\left( \begin{array}{cc}
		2i w_3 & i w_1 + w_2  \\
		i w_1 - w_2  & 0
	\end{array} \right)
=
	i w_3 I
	+ i w_1 \sigma_1
	+ i w_2 \sigma_2
	+ i w_3 \sigma_3,
\end{equation}
where $ \{ \sigma_j\} $ are the Pauli matrices. Its exponentiation is
\begin{equation}
	e^{tX} =
	e^{i t w_3}
	( I \cos \rho t
	+ i \vect{n} \cdot \vect{\sigma} \sin \rho t ),
\end{equation}
where $ \rho $ and $ \vect{n} $ are defined as
\begin{equation}
	\rho
	:= \| \vect{w} \| 
	= \sqrt{ (w_1)^2 + (w_2)^2 + (w_3)^2 },
	\qquad
	\vect{w} = \|\vect{w}\| \vect{n}.
\end{equation}
The associated horizontal extremal curve (\ref{extremal solution}) then becomes
\begin{equation}
	V(t) 
	= e^{t X} \, V_0 \, e^{-t {\mit\Omega}}
	= 
	e^{-i t w_3}
	\left(
		\begin{array}{l}
		\cos \rho t +i n_3 \sin \rho t \\
		( i n_1 - n_2)     \sin \rho t
		\end{array}
	\right)
	\label{extremal solution in N=2}
\end{equation}
and the projected curve in $ S^2 $ becomes
\begin{eqnarray}
	P(t)
&=&	V(t) V^\dagger (t)
	\nonumber \\
&=&
	\frac{1}{2} I +
	\frac{1}{2} \vect{\sigma} \cdot
	\left[
		\vect{n} (\vect{n} \cdot \vect{e}_3 )
		+
		( \vect{e}_3 - \vect{n} (\vect{n} \cdot \vect{e}_3 ) )
		\cos 2 \rho t
		-
		( \vect{n} \times \vect{e}_3 )
		\sin 2 \rho t
	\right],
	\label{projected curve}
\end{eqnarray}
where $ \vect{e}_3 = (0,0,1) $.
We see from (\ref{projected curve}) that
the point $ P(t) $ in $ S^2 $ starts at the north pole $ \vect{e}_3 $
of the sphere
and moves along a small circle with the axis $ \vect{n}$ in the clockwise 
sense by the angle $ 2 \rho t $.
The point $ P(t) $
comes back to the north pole when $ t $ satisfies $ 2 \rho t = 2 \pi n $
with an integer $ n $.
To make a closed loop, namely,
to satisfy the loop condition (\ref{loop condition}) at $ t = T = 1 $,
the control parameters must  satisfy
\begin{equation}
	\rho = \| \vect{w} \| = n \pi
	\qquad
	( n = \pm1, \pm2, \dots ).
\end{equation}
Then, the point $ P(t) $ travels the same small circle $ n $ times
during $ 0 \le t \le 1 $.
Therefore, the integer $ n $ counts the winding number of the loop.
At $t=1$, $ \cos \rho = (-1)^n $ and the holonomy
(\ref{holonomy condition}) is evaluated as
\begin{equation}
	V^\dagger_0 \, e^{X} \, V_0 \, e^{- {\mit\Omega}} 
	= e^{i w_3} (-1)^n e^{-2i w_3}
	= e^{-i (w_3 - n \pi)}
	= U_{\rm gate}
	= e^{i \gamma}.
	\label{Berry}
\end{equation}
Thus, to generate the holonomy $ U_{\rm gate} = e^{i \gamma} $,
the controller parameters are fixed as
\begin{equation}
	w_3 = n \pi - \gamma,
	\qquad
	w_1 +i w_2 = e^{-i \phi} \sqrt{ (n \pi)^2 - (n \pi - \gamma)^2 }.
\end{equation}
This is the solution to the inverse problem
defined by (\ref{loop condition}) and (\ref{holonomy condition}).
Here the nonvanishing integer $ n $ must satisfy
$ (n \pi)^2 - (n \pi - \gamma)^2 > 0 $.
The real parameter $ \phi $ is not fixed by
the loop condition and the holonomy condition.
The phase $ h_2 = e^{i \phi} $
parametrizes solutions in an equivalence class 
as observed in (\ref{equivalent X}).
The integer $ n $ classifies inequivalent classes.

The length of the loop, (\ref{length}), is now evaluated as
\begin{equation}
	S 
	= \tr ( W^\dagger W ) \, T
	= (n \pi)^2 - (n \pi - \gamma)^2.
\end{equation}
For a fixed $ \gamma $ in the range $ 0 \le \gamma < 2\pi $,
the simple loop with $ n = 1 $ is the shortest one among the extremal loops.
Thus, we conclude that the controller of $ U_{\rm gate} = e^{i \gamma} $ is
\begin{equation}
	X 
=
	\left( \begin{array}{cc}
		2i ( \pi - \gamma) & 
		i e^{ i \phi} \sqrt{ \pi^2 - (\pi - \gamma)^2 } \\
		i e^{-i \phi} \sqrt{ \pi^2 - (\pi - \gamma)^2 } & 0
	\end{array} \right).
	\label{Berry result}
\end{equation}
We call this solution a small circle solution
because of its geometric picture mentioned above.

\subsection{$U(k)$ holonomy}

Here we give a prescription to construct a controller matrix $ X $
that generates a specific unitary gate $ U_{\rm gate} $.
In other words, we give a systematic method to solve the inverse problem
(\ref{holonomy condition}).
It turns out that the working space should have a dimension
$ N \ge 2k $ to apply our method.
In the following we assume that $ N = 2k $.
The time interval is normalized as $ T = 1 $ as before.

Our method consists of three steps:
first, diagonalize the unitary matrix $ U_{\rm gate} $ to be implemented,
second, construct a diagonal controller matrix 
by combining small circle solutions,
third, undo diagonalization of the controller.

In the first step, 
we diagonalize a given unitary matrix $ U_{\rm gate} \in U(k) $ as
\begin{equation}
	R^\dagger U_{\rm gate} R 
	= U_{\rm diag}
	= \mbox{diag} ( e^{i \gamma_1}, \dots, e^{i \gamma_k} )
	\label{det0}
\end{equation}
with $ R \in U(k) $.
Each eigenvalue $ \gamma_j $ is taken in the range $ 0 \le \gamma_j < 2\pi $.
In the second step, we combine single loop solutions
associated with the Berry phase to construct two $ k \times k $ matrices
\begin{eqnarray}
&&	{\mit\Omega}_{\rm diag} 
	= \mbox{diag} ( i \omega_1, \dots, i \omega_k ),
	\qquad
	\omega_j = 2 ( \pi - \gamma_j),
	\label{det1}
	\\
&&	W_{\rm diag} 
	= \mbox{diag} ( i \tau_1, \dots, i \tau_k ),
	\qquad
	\tau_j = e^{i \phi_j} \sqrt{ \pi^2 - (\pi - \gamma_j)^2 }.
	\label{det2}
\end{eqnarray}
Then we obtain a diagonal controller
$$ 
	X_{\rm diag} = 
	\left( 
		\begin{array}{cc}
		{\mit\Omega}_{\rm diag} & W_{\rm diag}  \\
		- W^\dagger_{\rm diag} & 0
		\end{array}
	\right).
$$ 
In the third step, we construct the controller $ X $ as
\begin{equation}
	X 
	= 
	\left( 
		\begin{array}{cc}
		R & 0 \\
		0 & I_k
		\end{array}
	\right)
	\left( 
		\begin{array}{cc}
		{\mit\Omega}_{\rm diag} & W_{\rm diag}  \\
		- W^\dagger_{\rm diag} & 0
		\end{array}
	\right)
	\left( 
		\begin{array}{cc}
		R^\dagger & 0 \\
		0 & I_k
		\end{array}
	\right)
	= 
	\left( 
		\begin{array}{cc}
		R {\mit\Omega}_{\rm diag} R^\dagger & R W_{\rm diag}  \\
		- W^\dagger_{\rm diag} R^\dagger & 0
		\end{array}
	\right),
	\label{det3}
\end{equation}
which is a $ 2k \times 2k $ matrix.
We call
the set of equations,
(\ref{det0}), 
(\ref{det1}), 
(\ref{det2}) and
(\ref{det3}),
constructing equations of the controller.
This is the main result of this paper. 

It is easily verified that the controller $ X $ constructed above
satisfies the holonomy condition (\ref{holonomy condition}).
The diagonal controller $ X_{\rm diag} $ is actually a direct sum of 
controllers 
(\ref{Berry result}),
which generate Berry phases $ \{ e^{i \gamma_j} \} $.
Hence, its holonomy is also a direct sum of the Berry phases
(\ref{Berry}) as
$$ 
	V^\dagger_0 \, e^{X_{\rm diag}} \, V_0 \, e^{- {\mit\Omega}_{\rm diag} }
	= U_{\rm diag}
$$ 
and hence we have
$$ 
	V^\dagger_0 \, e^{X} \, V_0 \, e^{- {\mit\Omega}}
	= R 
	V^\dagger_0 \, e^{X_{\rm diag}} \, V_0 \, 
	R^\dagger R \,
	e^{- {\mit\Omega}_{\rm diag} }
	R^\dagger
	= R U_{\rm diag} R^\dagger
	= U_{\rm gate}.
$$ 

\section{Optimizing holonomic quantum computation}

Now we apply the prescription developed so far
to construct controllers
of several specific unitary gates, 
which are fundamental ingredients of quantum computation.
Our examples are 
the Hadamard gate,
the CNOT gate,
and
the two-qubit discrete Fourier transformation (DFT) gate.
For each unitary gate $ U_{\rm gate} $,
we need to calculate the diagonalizing matrix $ R $. 
Then the constructing equations of the controller,
(\ref{det0})-(\ref{det3}),
provide the desired optimal controller matrices.

\subsection{Hadamard gate}

The Hadamard gate is a one-qubit gate defined as
\begin{equation}
	U_{\rm Had} 
	= \frac{1}{\sqrt{2}} 
	\left( \begin{array}{cc}
		1&1\\
		1&-1
	\end{array} \right).
\end{equation}
It is diagonalized by
\begin{equation}
	R = 
	\left( \begin{array}{rr}
		\cos \frac{\pi}{8} & -\sin \frac{\pi}{8} \vspace{1mm} \\
		\sin \frac{\pi}{8} &  \cos \frac{\pi}{8}
	\end{array} \right)
\end{equation}
as
\begin{equation}
	R^\dagger U_{\rm Had} R = 
	\left( \begin{array}{cc}
		1 &  0 \\
		0 & -1 
	\end{array} \right).
\end{equation}
Needless to say,
\begin{equation}
	\cos \frac{\pi}{8} = \frac{\sqrt{ 2 + \sqrt{2} }}{2},
	\qquad
	\sin \frac{\pi}{8} = \frac{\sqrt{ 2 - \sqrt{2} }}{2}.
\end{equation}
Therefore, we have
$ \gamma_1 = 0 $ and $ \gamma_2 = \pi $.
We may put $ \phi_1 = \phi_2 = 0 $.
The ingredients of the constructing equations of the controller,
(\ref{det0})-(\ref{det3}), are calculated as
\begin{equation}
	{\mit\Omega}_{\rm diag} 
	= \mbox{diag} ( 2i \pi, 0 ),
	\qquad
	W_{\rm diag} 
	= \mbox{diag} ( 0, i \pi ),
\end{equation}
and hence
\begin{equation}
	R {\mit\Omega}_{\rm diag} R^\dagger =
	\frac{i \pi}{ \sqrt{2}} 
	\left(
		\begin{array}{cc}
		\sqrt{2}+1 & 1 \\
		1 & \sqrt{2}-1
		\end{array}
	\right),
	\qquad
	R W_{\rm diag} =
	\frac{i \pi}{2}
	\left( \begin{array}{rr}
		0 & - \sqrt{ 2 - \sqrt{2} } \vspace{1mm} \\
		0 &   \sqrt{ 2 + \sqrt{2} }
	\end{array} \right).
\end{equation}
Substituting these into (\ref{det3}), we obtain the optimal controller
of the Hadamard gate.

\subsection{CNOT gate}
One of the most important 2-qubit gates is the CNOT gate defined as
\begin{equation}
	U_{\rm CNOT} = 
	\left(\begin{array}{cccc}
		1&0&0&0 \\
		0&1&0&0 \\
		0&0&0&1 \\
		0&0&1&0
	\end{array} \right).
\end{equation}
It is diagonalized by
\begin{equation}
	R = \frac{1}{\sqrt{2}}
	\left(\begin{array}{cccc}
		\sqrt{2} &0&0&0 \\
		0& \sqrt{2} &0&0 \\
		0&0&1&-1 \\
		0&0&1&1
	\end{array} \right)
\end{equation}
as
\begin{equation}
	R^\dagger U_{\rm CNOT} R = 
	\left(\begin{array}{cccc}
		1&0&0&0 \\
		0&1&0&0 \\
		0&0&1&0 \\
		0&0&0&-1
	\end{array} \right).
\end{equation}
Therefore, we have
$ \gamma_1 = \gamma_2 = \gamma_3 = 0 $ and $ \gamma_4 = \pi $.
The ingredients of the controller are
\begin{equation}
	{\mit\Omega}_{\rm diag} 
	= \mbox{diag} ( 2i \pi, 2i \pi, 2i \pi, 0 ),
	\qquad
	W_{\rm diag} 
	= \mbox{diag} ( 0, 0, 0, i \pi ),
\end{equation}
and hence
\begin{equation}
	R {\mit\Omega}_{\rm diag} R^\dagger =
	i \pi
	\left(
		\begin{array}{cccc}
		2 & 0 & 0 & 0 \\
		0 & 2 & 0 & 0 \\
		0 & 0 & 1 & 1 \\
		0 & 0 & 1 & 1 
		\end{array}
	\right),
	\qquad
	R W_{\rm diag} =
	\frac{i \pi}{\sqrt{2}}
	\left(\begin{array}{cccc}
		0&0&0&0 \\
		0&0&0&0 \\
		0&0&0&-1 \\
		0&0&0&1
	\end{array} \right)
\end{equation}
Substituting these into (\ref{det3}), we obtain the optimal controller
of the CNOT gate.

\subsection{DFT2 gate}
Discrete Fourier transformation (DFT) gates are important 
in many quantum algorithms including Shor's algorithm for integer factorization.
The two-qubit DFT (DFT2) is a unitary transformation
\begin{equation}
	U_{{\rm DFT}2}
	= \frac{1}{2}\left( \begin{array}{cccc}
		1&1&1&1\\
		1&i&-1& -i\\
		1& -1& 1& -1\\
		1& -i& -1& i
	\end{array} \right).
\end{equation}
It is diagonalized by
\begin{equation}
	R = \frac{1}{2}
	\left(\begin{array}{cccc}
		1 & \sqrt{2} &-1 & 0 \\
		1 & 0        & 1 &-\sqrt{2} \\
		-1& \sqrt{2} & 1 & 0 \\
		1 & 0        & 1 & \sqrt{2} 
	\end{array} \right)
\end{equation}
as
\begin{equation}
	R^\dagger U_{{\rm DFT}2} R = 
	\left(\begin{array}{cccc}
		1&0&0&0 \\
		0&1&0&0 \\
		0&0&-1&0 \\
		0&0&0&i
	\end{array} \right).
\end{equation}
Therefore, we have
$ \gamma_1 = \gamma_2 = 0 $,
$ \gamma_3 = \pi $ and $ \gamma_4 = \pi/2 $.
Thus the ingredients of the controller are
\begin{equation}
	{\mit\Omega}_{\rm diag} 
	= \mbox{diag} ( 2i \pi, 2i \pi, 0, i \pi ),
	\qquad
	W_{\rm diag} 
	= \mbox{diag} ( 0, 0, i \pi, i \pi \sqrt{3}/2 ),
\end{equation}
and hence
\begin{equation}
	R {\mit\Omega}_{\rm diag} R^\dagger =
	\frac{i \pi}{2}
	\left(
		\begin{array}{cccc}
		3 & 1 & 1 & 1 \\
		1 & 2 &-1 & 0 \\
		1 &-1 & 3 &-1 \\
		1 & 0 &-1 & 2 
		\end{array}
	\right),
	\qquad
	R W_{\rm diag} =
	\frac{i \pi}{2}
	\left(\begin{array}{cccc}
		0 & 0 &-1 & 0 \\
		0 & 0 & 1 &-\sqrt{3/2} \\
		0 & 0 & 1 & 0 \\
		0 & 0 & 1 & \sqrt{3/2} 
	\end{array} \right).
\end{equation}
Substituting these into (\ref{det3}), we finally obtain the optimal controller
of the DFT2 gate.

\section{Summary and discussions}

Let us summarize our argument.
We briefly reviewed the WZ holonomy
and discussed that it may be utilizable for implementation
of quantum computation.
The WZ holonomy is neatly described in terms of differential geometry
of a homogeneous bundle, 
which consists of Stiefel and Grassmann manifolds
and is equipped with the canonical connection.
We formulated 
the optimization problem of control in holonomic quantum computation
in a form of the isoholonomic problem in the homogenous bundle.
We would like to emphasize that it had been left unsolved for
more than a decade after the first proposal.
We derived a set of equations,
(\ref{horizontal}) and (\ref{extremal}),
that characterizes the optimal control
and solved it to obtain the horizontal extremal curve (\ref{complete solution}).
The curve must satisfy two boundary conditions,
(\ref{loop condition}) and (\ref{holonomy condition}),
to be a closed loop in the control manifold and
to produce a specified unitary gate as a holonomy.
We solved this inverse problem 
by combining small circle solutions (\ref{Berry result}) to $U(1)$ holonomy
into a direct sum.
We provided a prescription (\ref{det0})-(\ref{det3})
to construct exactly an optimal controller for any unitary gate.
Finally we applied our prescription to
several important quantum gates.
transform gate.

We would like to discuss prospective development of the 
results presented above.
Although our prescription is applicable to 
arbitrarily large qubit gates,
the homogeneous bundle seems rather over-idealized for practical applications.
A realistic quantum system may have smaller control manifold $ M $
than the Grassmann manifold.
The restricted control manifold $ M $ is embedded in the Grassmann manifold
by an embedding map $ f : M \to G_{N,k} (\C) $ and
we need to study the isoholonomic problem
in the pullbacked bundle $ f^* S_{N,k} (\C) $.
Furthermore, the available working Hilbert space in a realistic system may 
not have dimensions as large as $ N \ge 2k $.
Actually, even when $ N <2k $,
sequential operations of single loop solutions can generate any unitary gate.
However, such a patched solution could not be optimal.
These problems will be treated separately in our future publications.



\section*{Acknowledgements}

ST is partially supported by Japan Society for the Promotion of Science (JSPS)
(Project No.~15540277).
MN would like to thank partial supports of Grant-in-Aids for 
Scientific Research from the Ministry of Education, 
Culture, Sports, Science and Technology, Japan (Project No.~13135215)
and from JSPS (Project No.~14540346).

\end{document}